\documentclass[12pt,a4paper]{article}
\begin{document}
\textwidth=135mm
 \textheight=200mm
\begin{center}
{\bfseries Towards Lagrangian formulations of mixed-symmetry
Higher Spin Fields  on AdS-space within BFV-BRST formalism
\footnote{{\small Talk on the International Bogolyubov
Conference-2009 "Problems of Theoretical and Mathematical
Physics", Moscow - Dubna, August 21 - 27, 2009.}}} \vskip 5mm A.A.
Reshetnyak\vskip 5mm {\small {\it $^\dag$ Institute of  Strength
Physics and Materials Science, 634021, Tomsk, Russia}}
\\
\end{center}
\vskip 5mm \centerline{\bf Abstract} The spectrum of superstring
theory on the $AdS_5 \times S_5$ Ramond-Ramond background in
tensionless limit contains integer and half-integer higher-spin
fields subject at most to  two-rows Young tableaux $Y(s_1,s_2)$.
We review the details of a gauge-invariant Lagrangian description
of such massive and massless higher-spin fields in anti-de-Sitter
spaces with arbitrary dimensions. The procedure is based on the
construction of Verma modules, its oscillator realizations and of
a BFV-BRST operator for non-linear algebras encoding unitary
irreducible representations of AdS group. \vskip 10mm
\section{\label{sec:intro}Introduction}
Launch of LHC on the rated capacity  assumes not only the answer
on the question on existence of Higgs boson,  the proof of
supersymmetry display and a new insight on origin of Dark Matter
\cite{LHC},
 but permits one to reconsider the problems of an
unique description of variety of elementary particles and all
known interactions. In this relation, the development of
higher-spin (HS) field theory in view of its close relation to
superstring theory on constant curvature spaces, which operates
with an infinite set of massive and massless bosonic and fermionic
HS fields subject to multi-row Young tableaux (YT)
$Y(s_1,...,s_k)$, $k \geq 1$ (see for a review, \cite{reviews})
seems by actual one. The paper considers the last results of
constructing
 Lagrangian formulations (LFs) for free integer and half-integer HS fields on
$AdS_d$-space with $Y(s_1,s_2)$ in Fronsdal metric-like formalism
within  BFV-BRST approach \cite{BFV} as a starting point for an
interacting HS field theory in the framework of conventional
Quantum Field Theory, and in part based on the results presented
in \cite{adsfermBKR,flatfermmix, 0812.2329, Sakharov}.

This method of Lorentz-covariant constructing LF for HS fields,
developed originally in a way that applies to Hamiltonian
quantization of gauge theories with a given LF, consists in a
solution of the \textit{problem inverse} to that of the method
\cite{BFV} (as in the case of string field theory \cite{SFT} and
in the early papers on HS fields \cite{Ouvry})  in the sense of
constructing a classical gauge LF with respect to a nilpotent
BFV--BRST operator $Q$.

In detail, the solution of \emph{inverse problem}  includes 4
steps:
\begin{itemize}\vspace{-0.5ex}
    \item  the realization of initial irrep conditions
of AdS group, that extract the fields with a definite mass $m$ and
generalized spin $\mathbf{s}=(s_1,...,s_k)$ \cite{Metsaev}  as
operator mixed-class constraints $o_I$ in a special Fock space
$\mathcal{H}$; \vspace{-1ex}
\item the additive conversion
(following to \cite{conversion}) of  algebra $o_I$ into one of
$O_I$: $O_I=o_I+o'_I$, $[o_I,o'_J\}=0$,  determined on wider Fock
space, $\mathcal{H}\bigotimes \mathcal{H}'$ with only first-class
constraints $O_\alpha \subset O_I$; \vspace{-1ex}
    \item the construction of the
  Hermitian nilpotent BFV-BRST operator $Q'$ for non-linear superalgebra of
 converted operators $O_I$ which  contains the  BFV-BRST operator $Q$ for only
 subsystem of $O_\alpha$;
\vspace{-1ex}
    \item the finding  of Lagrangian $\mathcal{L}$ for given HS field
     through corresponding scalar product $\langle \ | \
\rangle$  like $\mathcal{L} \sim \langle \chi |Q |\chi \rangle$,
to be invariant with respect to
 gauge transformations $\delta|\chi \rangle = Q
|\Lambda \rangle$ with $|\chi \rangle$ containing initial HS
field.
\end{itemize}
%The application of above algorithm  for bosonic \cite{flatbos} and
%fermionic \cite{flatfermmix, flatferm}
% HS fields on flat spaces  did not meet the
%problems in the most complicated second and third steps due to Lie
%(super)algebra  structure of initial constraints $o_I$: $[o_I,
%o_J\} = f_{IJ}^K o_K$. Indeed, for the same (super)algebra of
%additional parts $o'_I$ it is sufficient to use the construction
%of Verma module (VM)  for the superalgebra $osp(2k|k)$ (for
%integer spin algebra $sp(2k)$) \cite{simplealg} which is in
%one-to-one correspondence with unitary irreps of Lorentz algebra
%$so(1,d-1)$ subject to  $Y(s_1,...,s_k)$, $k\leq
%\left[\frac{d}{2}\right]$ due to Howe duality. Then an oscillator
%realization of the VM  in Fock space $\mathcal{H}'$ represents a
%polynomial form as compared to totally-symmetric HS fields on
%$AdS_d$ space \cite{adsfermBKR, 0206027} where, (super)algebras of
%$o_I$ are non-linear, not coinciding with  ones for $o'_I$ due to
%presence of inverse squared $AdS_d$-radius $r$ ($r = R/d(d-1)$ for
%scalar curvature $R$) \cite{0905.2705, 0206027}. , for coordinate
%$\mathcal{C}^I$ and
%    momenta $\mathcal{P}_J$ ghosts.
\vspace{-1.0ex} As compared to application of above algorithm  for
bosonic \cite{flatbos} and fermionic \cite{flatfermmix, flatferm}
 HS fields on $\mathbf{R}^{1,d-1}$  with standard resolution of the
 2nd and 3rd steps  due to the same Lie
 (super)algebra structure for $o_I, o'_I, O_I$: $[o_I,
o_J\} = f_{IJ}^K o_K$, their resolution already for
totally-symmetric HS fields on $AdS_d$ space \cite{adsfermBKR,
0206027} are not so easy. It is revealed on stages of Verma module
(VM) construction for $o'_I$ and its non-polynomial(!) oscillator
realization in $\mathcal{H}'$  because of  $AdS$-radius
$r^{-\frac{1}{2}}$ ($r = R/d(d-1)$ for scalar curvature $R$)
\cite{adsfermBKR, 0206027}. In turn, a construction of BFV-BRST
operator $Q'$ does not have the Lie-algebra form, ${Q'} =
\mathcal{C}^I {O}_I  + \textstyle\frac{1}{2}
\mathcal{C}^I\mathcal{C}^{J}f^{K}_{JI}\mathcal{P}_K$   for
(super)algebra of
 $O_I$  in transiting to $AdS$-space.
%: ${Q} = \mathcal{C}^I {O}_I  +
%\textstyle\frac{1}{2}    \mathcal{C}^I\mathcal{C}^{J}f^{K}_{JI}\mathcal{P}_K$.

The main goals of the paper are to apply the above strategy to
construct   LFs for bosonic and fermionic HS fields on
$AdS_d$-spaces subject to $Y(s_1,s_2)$.
%The paper is organized as follows. In Section~2, we examine the
%initial operator algebra $(\mathcal{A})\mathcal{A}_{b}$. In
%Section~3, we consider Proposition, which determines a way to
%obtain algebraic relations for the (super)algebras of the parts
%$(\mathcal{A}')\mathcal{A}'_{b}$ additional to those for a
%specially modified (super)algebra
%$(\mathcal{A}_{mod})\mathcal{A}_{b{}mod}$, and examine a
%construction of Verma modules that realize the highest-weight
%representation of $(\mathcal{A}')\mathcal{A}'_{b}$ and their
%realization in an auxiliary Fock space. An exact BFV--BRST
%operator for a converted (super)algebra
%$(\mathcal{A}_{c})\mathcal{A}_{b{}c}$ is obtained in Section~4, on
%the basis of a solution of the Jacobi identity, due to the absence
%of non-trivial higher-order relations for
%$(\mathcal{A}_{c})\mathcal{A}_{b{}c}$. The action and the sequence
%of reducible gauge transformations, mainly for bosonic HS fields
%of a fixed spin $\mathbf{s}=(s_1,s_2)$, are deduced in Section~5.
%In the conclusion, we summarize the results of this article and
%discuss some open problems.
%
%We mainly use the conventions of
%Refs.~\cite{flatfermmix,adsfermBKR}.
\section{Bosonic  fields in AdS spaces}
A massive integer spin $\mathbf{s}=(s_1,s_2)$, ($s_1 \geq s_2$),
representation of the AdS group in an $AdS_d$ space is realized in
a space of mixed-symmetry tensors,
\begin{equation}\label{Young k2}
\Phi_{(\mu)_{s_1},(\nu)_{s_2}} \hspace{-0.2em}\equiv
\hspace{-0.2em}
\Phi_{\mu_1\ldots\mu_{s_1},\nu_1\ldots\nu_{s_2}}(x)
\hspace{-0.3em}\longleftrightarrow \hspace{-0.3em}
\begin{array}{|c|c|c c c|c|c|c|c|c| c|}\hline%\vphantom{\biggm|}
  \!\mu_1 \!&\! \mu_2\! & \cdot \ & \cdot \ & \cdot \ & \cdot\  & \cdot\  & \cdot\ &
  \cdot\    &\!\! \mu_{s_1}\!\! \\
   \hline%\vphantom{\biggm|}
    \! \nu_1\! &\! \nu_2\! & \cdot\
   & \cdot\ & \cdot  & \cdot &  \cdot & \!\!\nu_{s_2}\!\!   \\
  \cline{1-8}%\vphantom{\biggm|}
\end{array}\ ,
\end{equation}
subject to the Klein-Gordon (\ref{Eq-0b}), divergentless,
traceless and mixed-symmetry equations (\ref{Eq-1b}) [for $\beta =
(2;3) \Longleftrightarrow (s_1>s_2; s_1 = s_2)$]:
\begin{eqnarray}
\label{Eq-0b} &&\bigl[\nabla^2 +r[(s_1-\beta-1+ d)(s_1-\beta) -
s_1- s_2]+m^2 \bigr]\Phi_{(\mu)_{s_1},\ (\nu)_{s_2}} =0,\\
&&\bigl(\nabla^{\mu_1}, \ \nabla^{\nu_1}, \ g^{\mu_1\mu_2} ,\
g^{\nu_1\nu_2},\ g^{\mu_1\nu_1} \bigr)\Phi_{(\mu)_{s_1},\
(\nu)_{s_2}} = \Phi_{\{(\mu)_{s_1},\nu_1\}\nu_2...\nu_{s_2}}=0.
\label{Eq-1b}
\end{eqnarray}
To obtain HS symmetry algebra (of $o_I$)  for a
 description of all integer HS fields, we introduce a Fock
space $\mathcal{H}$, generated by 2 pairs of creation $a^i_\mu(x)$
and annihilation $a^{j+}_\mu(x)$ operators, $i,j =1,2, \mu,\nu
=0,1...,d-1$: $[a^i_\mu, a_\nu^{j+}]=-g_{\mu\nu}\delta_{ij}$, and
a set of constraints for an arbitrary string-like vector
$|\Phi\rangle \in \mathcal{H}$,
\begin{eqnarray}
\label{PhysState}  \hspace{-2ex}&& \hspace{-2ex} |\Phi\rangle  =
\textstyle\sum_{s_1=0}^{\infty}\sum_{s_2=0}^{s_1}\Phi_{(\mu)_{s_1},(\nu)_{s_2}}(x)\,
a^{+\mu_1}_1\ldots\,a^{+\mu_{s_1}}_1a^{+\nu_1}_2\ldots\,a^{+\nu_{s_2}}_2|0\rangle,\\
\label{l0} \hspace{-2ex}&& \hspace{-3ex} {\tilde{l}}_0|\Phi\rangle
= \bigl(l_0+ \tilde{m}^2_b + r \bigl((g_0^1-2\beta-2)g_0^1 -
 g_0^2 \bigr)\bigr)|\Phi\rangle=0 , \ l_0 = [D^2 -
r\textstyle\frac{d(d-6)}{4}],\\
\label{lilijt} \hspace{-2ex} && \hspace{-2ex} \bigl({l}^i, l^{ij},
t \bigr)|\Phi\rangle  = \bigl(-i a^i_\mu D^\mu,
\textstyle\frac{1}{2}a^{i}_\mu a^{j\mu}, a^{1+}_\mu a^{2\mu}\bigr)
|\Phi\rangle=0,\ i\leq j,
\end{eqnarray}
with number particles operators, $g_0^i =
-\frac{1}{2}\{a^{i+}_\mu,  a^{\mu{}i}\}$, central charge
$\tilde{m}^2_b = {m}^2$ + $r\beta(\beta+1)$, operator $D_\mu =
\partial_\mu-\omega_\mu^{ab}(x)\bigl(\sum_{i}a_{i{}a}^+a_{i{}b}
\bigr)$, $a^{(+)\mu}_i(x)=e^\mu_a(x)a^{(+)a}_i$ with  vielbein
$e^\mu_a$,  spin connection $\omega_\mu^{ab}$, tangent indices
$a,b$. Operator $D_\mu$  is equivalent in its action in
$\mathcal{H}$ to the covariant derivative $\nabla_{\mu}$ [with
d'Alambertian $D^2 = (D_a + \omega^{b}{}_{ba} )D^a$]. The set of 7
primary constraints (\ref{l0}), (\ref{lilijt}) with $\{o_\alpha\}$
= $\bigl\{{\tilde{l}}_0, {l}^i, l^{ij}, t \bigr\}$  are equivalent
to Eqs. (\ref{Eq-0b}), (\ref{Eq-1b}) for all spins.

For Hermiticity of BFV-BRST operator (reality Lagrangian
$\mathcal{L}$) the algebra with $o_\alpha$ must be enlarged by
adding the operators $\bigl(l^+_i, l^+_{ij}, t^+ \bigr)$,
resulting the HS symmetry algebra in AdS${}_d$ space with
$Y(s_1,s_2)$, denoted as $A(Y(2), AdS_d))$. The maximal Lie
subalgebra  of operators $l_{ij}, t, g_0^i, l^+_{ij}, t^+$ is
isomorphic to $sp(4)$  whereas the only nontrivial quadratic
commutators in $A(Y(2), AdS_d))$ are due to operators with
$D_{\mu}$: $l_i, \tilde{l}_0, l^+_i$. For the aim of LF
construction it is enough to have a simpler, (so called
\textit{modified}) algebra $A_{mod}(Y(2), AdS_d))$, with operator
$l_0$ (\ref{l0}) instead of ${\tilde{l}}_0$, so that
 AdS-mass term, $\tilde{m}^2_b + r
\bigl((g_0^1-2\beta-2)g_0^1 -
 g_0^2 \bigr)$, will be restored later within conversion
 and  properly construction of LF.
 Algebra $A_{mod}(Y(2), AdS_d))$
contains 1 first-class constraint $l_0$,  4 differential $l_i,
l_i^+ $,  8 algebraic $ t, t^+, l_{ij}, l_{ij}^+$ second-class
constraints $\theta_{\mathbf{a}}$, operators $g_0^i$, composing an
invertible matrix: $\|[\theta_{\mathbf{a}},
\theta_{\mathbf{b}}\}\| = \|\Delta_{\mathbf{ab}}(g_0^i)\| +
(o_{I})$, and satisfies the non-linear relations (additional to
ones for $sp(4)$)  given by Table~\ref{table}.
\hspace{-1ex}\begin{table}[t] {
\begin{center}
\begin{tabular}{||c||c|c|c|c|c|c|c||c||}\hline\hline
$\hspace{-0.2em}[\; \downarrow, \rightarrow \}$\hspace{-0.7em}&
 $t$ & $t^+$ & $l_0$ &
$l^i$ &$l^{i{}+}$ & $l^{ij}$
&$l^{ij{}+}$ &$g^i_0$ \\
\hline\hline $l_0$
    & $0$ & $0$ & $0$
   & \hspace{-0.3em}
    $\hspace{-0.2em}-r{\mathcal{K}}^{bi+}_1$\hspace{-0.5em} & \hspace{-0.3em}
    $r{\mathcal{K}}^{bi}_1$\hspace{-0.3em}
    & $0$ & $0$ & $0$ \\
\hline $l^k$
   & \hspace{-0.5em}$- l^2\delta^{k1}$ \hspace{-0.5em} & \hspace{-0.5em}$
   -l^1\delta^{k2}$ \hspace{-0.9em}&\hspace{-0.9em}
   $r{\mathcal{K}}^{bk+}_1$\hspace{-1.1em}
   & \hspace{-0.3em}${W}^{ki}_b$ \hspace{-0.3em} & \hspace{-0.3em}
   ${X}^{ki}_b$\hspace{-0.3em}
    & $0$ & \hspace{-0.5em}$- \textstyle\frac{1}{2}l^{\{i+}\delta^{j\}k}$
    \hspace{-0.9em} & \hspace{-0.4em}
    $l^i\delta^{ik}$\hspace{-0.4em} \\
\hline $l^{k+}$ & \hspace{-0.5em}$l^{1+}
   \delta^{k2}$\hspace{-0.7em} & \hspace{-0.7em}$l^{2+}\delta^{k1}$ \hspace{-1.0em} &
   \hspace{-0.5em}$-r{\mathcal{K}}^{bk}_1$ \hspace{-1.1em}&\hspace{-0.3em}
   $-{X}^{ik}_b$\hspace{-0.3em}
   &\hspace{-0.5em} $- {W}^{ki+}_b$\hspace{-0.5em}
    &\hspace{-0.7em} $ \textstyle\frac{1}{2}l^{\{i}\delta^{j\}k}$\hspace{-0.7em} & $0$ &\hspace{-0.7em} $-l^{i+}\delta^{ik}$\hspace{-0.5em}  \\
   \hline\hline
\end{tabular}
\end{center}} \vspace{-2ex}\caption{The  non-linear part of algebra $A_{mod}(Y(2),
AdS_d)$.}\label{table}
\end{table}

%\noindent
 In the Table~\ref{table},  the quantities ${\mathcal{K}}^{bk}_1, {W}^{ki}_b,
X^{ki}_b$ are quadratic in $o_I$ (see \cite{0812.2329} for
details)
\begin{eqnarray}
{} {W}^{ki}_b & = & 2r\varepsilon^{ki}\left[(g_0^2-g_0^1)l^{12} -
t l^{11} + t^+ l^{22}\right], \quad
\varepsilon^{ki}=-\varepsilon^{ik}, \varepsilon^{12}=1,
 \label{lilj}\\
{\mathcal{K}}^{bk}_1& = &\Bigl(4\textstyle\sum_{i}l^{ki+}l^i
+l^{k+}(2g_0^k-1) -2l^{2+}t \delta^{k1} - 2l^{1+}t^+
\delta^{k2}\Bigr)  \label{l'0li+} ,
\\
 {} {X}^{ik}_b
  & =
 &\bigl\{{{l}}_{0}+ r\bigl( K^{0i}_0  +
\mathcal{K}^{12}_0 \bigr)\bigr\}\delta^{ij}  + r\bigl\{
\bigl[4\textstyle\sum\nolimits_{j} l^{ 1j+}l^{ j2} +
\textstyle(\sum_{j}g_0^{ j}
  -{2}
)t\bigr]\delta^{k1}\delta^{i2}\nonumber\\
&& + r\bigl\{ \bigl[4\textstyle\sum\nolimits_{j} l^{ j2+}l^{ 1j} +
t^+\textstyle(\sum_{j}g_0^{ j}
  -{2}
)\bigr]\delta^{k2}\delta^{i1}, \label{lilj+b}
\end{eqnarray} with  $K^{0i}_0, i=1,2$, $\mathcal{K}^{12}_0$ composing
a Casimir operator $\mathcal{K}$ for $sp(4)$ algebra
\begin{equation}\label{Casimirs}
\mathcal{K}_0 =
   \textstyle\sum_{i}K_0^{0i} +
    2\mathcal{K}_0^{12} =
    \sum_{i}\bigl((g_0^i)^2-2g_0^i -4l^{ii+}l^{ii}\bigr)+2 \bigl(t^{ +}t - g_0^{ 2} -
    4l^{ 12 +}
      l^{ 12}\bigr).
\end{equation}
Algebra $A(Y(2), AdS_d)$ generalizes its analogs both for
mixed-symmetry  HS fields  on flat spaces and for
totally-symmetric HS fields  on AdS spaces \cite{flatbos,
0206027}.
\section{Verma module, Fock space realization}
The procedure of additive conversion for non-linear algebra
$A(Y(2), AdS_d)$ of $o_I$ into algebra $A_c(Y(2), AdS_d)$ of
converted constraints $O_I$, $O_I=o_I+o'_I$, $[o_I, o'_J\}$ acting
in Hilbert space $\mathcal{H}\bigotimes \mathcal{H}'$ with only
first-class constraints $O_\alpha$ \cite{0206027, adsfermBKR} may
be realized in two ways resulting either to unconstrained or to
constrained LF. For the former case, it means the conversion of
the total set of second-class constraints $\{\theta_\mathbf{a}\}$
for the latter the conversion of only differential and part of
algebraic constraints: $l_i, l_i^+, t, t^+$ having restricting the
algebra $\mathcal{A}(Y(2),AdS_d)$ to the  reduced non-linear
algebra $\mathcal{A}_r(Y(2),AdS_d)$ = $\{l_0, g_0^i, l_i, l_i^+,
t, t^+\}$ with off-shell traceless conditions  on the fields and
gauge parameters of final LF. For the aims of QFT an unconstrained
LF is more preferable.

To find  additional parts $o'_I$  we, first, determine the
multiplication law for the algebra $\mathcal{A}'(Y(2),AdS_d)$,
which for compactly written multiplication law for $o_I$ given by
Table~\ref{table} reads (see Refs.\cite{adsfermBKR, 0812.2329,
0206027} for details)
\begin{eqnarray}
[\,o_I',o_J'\}  = f_{IJ}^Ko_K'- f_{IJ}^{KM}o_M'o_K'
  \quad \mathrm{if} \quad [{o}_I,{o}_J\} =
f_{IJ}^K{o}_K+f_{IJ}^{KM}{o}_K{o}_M. \label{addal}
\end{eqnarray}
Second, following generalization of Poincare--Birkhoff--Witt
theorem, we construct VM, based on Cartan-like decomposition
enlarged from one for $sp(4)$
\begin{equation}\label{Cartandecomp}
    \mathcal{A}'(Y(2),AdS_d) =  \{l^{\prime +}_{ij},
t^{\prime+}, l^{\prime +}_i\} \oplus \{g_0^{\prime i}, l_0'\}
\oplus \{l^{\prime }_{ij}, t', l^{\prime }_i\} \equiv
\mathcal{E}^-\oplus H \oplus\mathcal{E}^+.
\end{equation}
Note, that in contrast to the case of Lie (super)algebra and
totally-symmetric HS fields on AdS space \cite{flatfermmix,
adsfermBKR, 0206027}, the negative root vectors $l^{\prime +}_1,
t^{\prime+}, l^{\prime +}_2$ are not commuted, making the
 arbitrary vector $|\vec{N}\rangle_V = \left| {n}_{11},{n}_{12}, {n}_{22},n_1, n, n_2
\rangle_V \right.$
\begin{equation}\label{VM}
   |\vec{N}\rangle_V \equiv  \textstyle\bigl(l^{\prime
+}_{11}\bigr){}^{n_{11}}\bigl(l^{\prime +}_{12}\bigr){}^{
n_{12}}\bigl(l^{\prime +}_{22}\bigr){}^{ n_{22}}
\Bigl(\frac{l^{\prime +}_1}{m_1}\Bigr){}^{ n_1}\bigl(t^{\prime
+}\bigr){}^{n} \Bigl(\frac{l^{\prime
+}_{2}}{m_2}\Bigr){}^{n_{2}}|0\rangle_V, \
\mathcal{E}^+|0\rangle_V=0
\end{equation}
 from VM, (for highest weight vector
$|0\rangle_V$,  $n_{ij}, n_i, n \in \mathbf{N}_0$,  and arbitrary
constants $m_i$ with dimension of mass) by not proper one for
$t^{\prime+}, l^{\prime
 +}_2$!
That \emph{nontrivial entanglement} are resolved within iterative
procedure, so that the VM for  algebra $\mathcal{A}'(Y(2),AdS_d)$
may be constructed.   Omitting tedious technical details (see
Ref.\cite{Sakharov}), note that by the crucial difference here is
the presence of so-called \emph{basic block} operator $\hat{t}'$
enlarging Lie part ${t}'_{L}$ in ${t}'={t}'_{L} + \hat{t}'$:
$\hat{t}'_{\vert r=0} =0$,
 from which the result of the action of
all $o'_I$ on vector $|\vec{N}\rangle_V$ is found. The realization
of the VM in formal power series (due to $r$) in degrees of
creation and annihilation operators $(B,B^+) = b_i, b_i^+, b_{ij},
b_{ij}^+, b, b^+$ in  $\mathcal{H}'$ whose number coincides to
ones of 2nd-class constraints  is solved too. As a result, the
Cartan generators have the boundary conditions: $(g_0^{\prime i},
l_0')=(h^i, m_0^2) + O(B,B^+)$, with constants $h^i, m_0^2$
permitting then under special choices to restore the correct
equations of motion within LF. The explicit form of basic block
$\hat{t}'(B,B^+)$ (in ${t}'={t}'_{L} + \hat{t}'$) reads,
{\small\begin{eqnarray}
   && \hspace{-1.0em}{t}'_{L} =\left(h^1-h^2 - b^+_2b_2 - b^+b\right)b - b_{11}^+b_{12}
   -2b_{12}^+b_{22}\,,
 \label{t'Lf}  \\
   && \hspace{-1.0em}\hat{t}' = \sum_{k=0}\Biggl[\sum_{{}^1m=0}\hspace{-0.2em}
   \sum_{{}^1l=0} ... \sum_{{}^km=0}\hspace{-0.2em}\sum_{{}^kl=0}
   (-1)^k\hspace{-0.2em}
   \left(\hspace{-0.2em}\frac{-2r}{m_2^2}\hspace{-0.2em}\right)^{\hspace{-0.2em}\sum_{i=1}^k({}^im+{}^il)+k}
   \hspace{-0.2em}\prod_{i=1}^k
   \frac{1}{(2{}^im+1)!}\frac{1}{(2{}^il+1)!}\nonumber
     \\
  \hspace{-1.0em} && \hspace{1.0em}\times \bigl(b_{22}^+\bigr)^{\sum_{i=1}^k({}^im+{}^il)+k}\Biggl\{
   b_2^+bb_2 -\frac{m_1}{m_2}\sum_{m=0}\left(\frac{-2r}{m_2^2}\right)^m
    \frac{(b_{22}^+)^m}{(2m+1)!}b_1^+
    b_2^{2m+1}\nonumber \\
  \hspace{-1.0em} &&\hspace{1.0em} +\sum_{m=1}\left(\frac{-2r}{m_2^2}\right)^m
   (b_{22}^+)^{m-1}\Bigl[b_{12}^+
   \Bigl\{
    \frac{h^2-h^1+2b^+b}{(2m)!}+\frac{b^+_2b_2}{(2m+1)!}\Bigr\}
   \nonumber \\
\hspace{-1.0em}   &&  \hspace{1.0em}- \frac{b_{11}^+b^+}{(2m)!} - \frac{b_{22}^+}{(2m)!}(h^2-h^1+b^+b)b\Bigr]b^{2m}_2 \nonumber \\
   \hspace{-1.0em}   &&  \hspace{1.0em} - \sum_{m=0}\sum_{l=0}\left(\frac{-2r}{m_2^2}\right)^{m+l+1}
   \frac{1}{(2m+1)!}(b_{22}^+)^{m+l}\Bigl[ b_{12}^+
   \Bigl\{\frac{h^2-h^1+2b^+b+b_2^+b_2}{(2l+1)!}\nonumber\\
 \hspace{-1.0em}   &&  \hspace{1.0em} -
   \frac{b_2^+b_2}{(2l+2)!}
   \Bigr\}
   -\frac{b_{11}^+b^+}{(2l+1)!}  -\frac{b_{22}^+}{(2l+1)!}(h^2-h^1+b_2^+b_2+b^+b)b \nonumber\\
   \hspace{-1.0em}   &&  \hspace{1.0em}+
   \frac{m_1}{m_2}\frac{b_{22}^+}{(2l+2)!}b_1^+b_2
   \Bigr]  b_2^{2(m+l+1)}
   \Biggr\}(b_2)^{2(\sum_{i=1}^k({}^im+{}^il)+k)}\Biggr],
\end{eqnarray}}
 whereas   the rest operators
$o'_I(B,B^+)$ may be found in \cite{Sakharov}.

Therefore, the solution of the second problem on construction of
the VM for algebra $A'(Y(2), AdS_d)$ and its oscillator
realization  is found.
%%%%%%%%%%%%%
\section{BRST operator for  non-linear algebra}
%%%%%%%%%%%%%%%%
The system of  ${O}_I$ forming  non-linear algebra
$\mathcal{A}_{c}(Y(2), AdS_d)$ with multiplication law following
from Eq. (\ref{addal}):
\begin{equation}\label{multAc} [O_I, O_J\} =
F_{IJ}^K(O)O_K,\quad F_{IJ}^K(O) = f_{IJ}^K
      -\bigl(f_{IJ}^{MK}+
      f_{IJ}^{KM}\bigr)o_M' + f_{IJ}^{MK}{O}_M
\end{equation} and
      Table~\ref{table} now has no the form of closed algebra,
      because of presence nontrivial Jacobi identities
for 6 triples $(L_1, L_2, L_0)$, $(L_1^+, L_2^+, L_0)$, $(L_i,
L_j^+, L_0)$. Indeed, there exists a set of third order structural
functions  in terminology of Ref. \cite{BFV} resolving those
identities (see \cite{0812.2329}).
 The construction of a BFV-BRST operator ${Q}'$ for  $\mathcal{A}_{c}(Y(2),
AdS_d)$,   in  case of the Weyl ordering for quadratic
combinations of ${O}_I$ in the r.h.s. of $[{O}_I, {O}_J\}$ and for
the $(\mathcal{C}\mathcal{P})$-ordering for the ghost coordinates
  $\mathcal{C}^I$: $\eta_0$,
$\eta^i_G$ $\eta_i^+$, $\eta_i$, $\eta_{ij}^+$, $\eta_{ij}$,
$\eta$, $\eta^+$,  and their conjugated momenta $\mathcal{P}_I$:
${\cal{}P}_0$, ${\cal{}P}^i_G$ ,  ${\cal{}P}_i$, ${\cal{}P}_i^+$,
${\cal{}P}_{ij}$, ${\cal{}P}_{ij}^+$, $\mathcal{P}^+$,
$\mathcal{P}$, is a more complicated task than for
totally-symmetric HS fields on AdS${}_d$ \cite{adsfermBKR,
0206027}.

The nilpotent operator ${Q}'$  has the terms in third degree in
ghosts $C^I$  \cite{0812.2329},
 \begin{eqnarray}\label{explQ'}
    &&\hspace{-1.5em} {Q'}
 =  Q'_1 + Q'_2 + \Bigl[r^2\eta_0\eta_i\eta_j\varepsilon^{ij}\Bigl\{
\textstyle \frac{1}{2}\sum_k\Bigl(G^k_0[\mathcal{P}
\mathcal{P}^{+}_{22}
 - \mathcal{P}^+
\mathcal{P}^{+}_{11}+ {i}
\mathcal{P}^{+}_{12}\sum_l(-1)^l\mathcal{P}_G^l]
   \nonumber\\
&& \textstyle -
  {i}({L}^{+}_{11}\mathcal{P}^+-{L}^{+}_{22}\mathcal{P})
  \mathcal{P}_G^k + 4L^{kk}\mathcal{P}^{+}_{k2}
\mathcal{P}^{+}_{1k}\Bigr)
 - {L}^{+}_{12}\mathcal{P}_G^1\mathcal{P}_G^2  +2
L^{12}\mathcal{P}^{+}_{22}\mathcal{P}^{+}_{11}\Bigr\}
\nonumber\\
&& \hspace{-0.7em}\textstyle + r^2\eta_0\eta^+_i\eta_j\Bigl\{
\Bigl[\sum_{k}(-1)^k\frac{i}{2}G^k_0 \sum_l\mathcal{P}_G^l +
2(L^{+}_{22}\mathcal{P}^{22}-L^{11}\mathcal{P}^{+}_{11})\Bigr]
\mathcal{P}\delta^{1j} \delta^{2i} \nonumber\\
&& \hspace{-0.7em}+
\varepsilon^{\{1j}\delta^{2\}i}\Bigl(\imath\textstyle\sum_k\Bigl[
\frac{1}{2}T\mathcal{P}^+ - 2 L^{12}\mathcal{P}^{+}_{12}
(-1)^k\Bigr]\mathcal{P}_G^k  +2(
L^{12}\mathcal{P}-T\mathcal{P}^{12})\mathcal{P}^{+}_{22}
\nonumber\\
&& \hspace{-0.7em} +
2(T^+\mathcal{P}^{12}-L^{12}\mathcal{P}^+)\mathcal{P}^{+}_{11}
\Bigr) -T\Bigl[\mathcal{P}_G^1\mathcal{P}_G^2\delta^{2i}
\delta^{1j}+2\mathcal{P}^{11}\mathcal{P}^{+}_{22}\delta^{1i}\delta^{2j}\Bigr]
- \textstyle
2\sum_{k}(-1)^k\nonumber\\
&& \hspace{-0.7em}
\times\Bigl[(G^k_0\mathcal{P}^{11}+\imath{L}^{11}\mathcal{P}_G^k)
\delta^{1i}\delta^{2j}- (G^k_0\mathcal{P}^{22} +
\imath{L}^{22}\mathcal{P}_G^k) \delta^{2i}\delta^{1j} \Bigr]
\mathcal{P}^{+}_{12}
  \Bigr\} + h. c.\Bigr], \end{eqnarray}
with the standard form for linear $Q_1'$ and quadratic $Q_2'$
terms in ghosts $\mathcal{C}^I$. The Hermiticity of operator $Q'$
is defined by the rule: $
  Q^{\prime +}K = K Q'$, {for} operator $K = \hat{1} \otimes K' \otimes
  \hat{1}_{gh}$, with non-degenerate operator $K'$ providing the Hermiticity of
the additional parts $o'_I$ in $\mathcal{H}'$.
\section{Unconstrained Lagrangian formulation}
 BFV-BRST operator $Q$ for 1st-class constraints
 $O_\alpha = \{L_0,L^i,L^{ij},T\}$ is extracted from $Q'$
 (\ref{explQ'}) by collecting the terms with ghosts $\eta^i_{G}$
 being by BRST-invariant
 enlarged number particle operators $\sigma^i + h^i = G_0^i  + ghosts$ and
 inessential operators $\mathcal{B}^i$ for further derivation of an LF,
\begin{eqnarray}
{Q}'  =  {Q} + \eta^i_G(\sigma^i+h^i) + \mathcal{B}^i
\mathcal{P}^i_G.     \label{decomposQ'}
\end{eqnarray}
The same is applied to a physical vector $|\chi\rangle \in
\mathcal{H}_{tot}=\mathcal{H}\bigotimes
\mathcal{H}'\bigotimes\mathcal{H}_{gh}$,
$|\chi\rangle$=$|\Phi\rangle + |\Phi_A\rangle$,
$|\Phi_A\rangle_{\vert \mathcal{H}}$ = $0$, with $|\Phi\rangle$
given in (\ref{PhysState}). From commutativity,  $[Q,
\sigma^k\}=0$ and choice of a representation for Hilbert space (as
in SFT \cite{SFT})) it follows the spectral problem from  the
equation ${Q}'|\chi\rangle = 0$ \cite{0812.2329},
\begin{equation}
\label{Qchi}  {Q}|\chi\rangle = 0, \quad
(\sigma^i+h^i)|\chi\rangle=0, \quad {gh}(|\chi\rangle)=0,
\end{equation}
thus determining the spectrum of spin values  and proper
eigenvectors,
\begin{eqnarray}
h^i(s_i)&=& \textstyle -\bigl( s_i + \frac{d - 5 }{2}
-2\delta^{i2} \bigr)\,, \label{hchin1n2}  \quad
|\chi\rangle_{(s_1,s_2)},
\end{eqnarray}
As differed to the second equation in (\ref{Qchi}) the first
equation is valid only in the subspace of $\mathcal{H}_{tot}$ with
the zero ghost number.

After substitution: $h^i \to   h^i(s_i)$ operator $Q_{(s_1,s_2)}$,
is nilpotent on each subspace $H_{tot{}(s_1,s_2)}$ whose vectors
satisfy to the Eqs.(\ref{Qchi}) for  (\ref{hchin1n2}). Hence, the
equations of motion (one to one correspond to
Eqs.(\ref{Eq-0b}),(\ref{Eq-1b})),~a~sequ\-ence of reducible gauge
transformations and Lagrangian action have the form
\begin{eqnarray}
\hspace{-1em}&& Q_{(s_1,s_2)}|\chi^0\rangle_{(s_1,s_2)}=0,  \
\delta|\chi^{l} \rangle_{(s_1,s_2)}
=Q_{(s_1,s_2)}|\chi^{1+1}\rangle_{(s_1,s_2)},
\ l = 0,...,6,\label{LEoM} \\
%%%%%%%%%%%%%%%%%%
\hspace{-1em}&& {\cal S}_{(s_1,s_2)} = \int d \eta_0 \;
{}_{(s_1,s_2)}\langle \chi^0 |K_{(s_1,s_2)} Q_{(s_1,s_2)}| \chi^0
\rangle_{(s_1,s_2)},\texttt{ for } |\chi^0\rangle\equiv
|\chi\rangle. \label{S}
\end{eqnarray}
The corresponding LF for bosonic field with spin $\mathbf{s}$
subject to $Y(s_1,s_2)$ is a reducible gauge theory of maximally
$L = 6$-th stage of reducibility.

\section{Fermionic  fields in AdS spaces} The
construction of LF for half-integer HS fields subject to
$Y(s_1,s_2)$, $s_i=n_i+\frac{1}{2}$ is more complicated due to
presence of fermionic constraints, which follows from respective
irrep conditions on spin-tensor $\Phi_{(\mu)_{n_1},\ (\nu)_{n_2}}$
(with suppressed spinor index $A$) being by Dirac, gamma-traceless
and mixed-symmetry equations
\begin{eqnarray}
 \Bigl(\bigl[i\gamma^{\mu}\nabla_{\mu}
    -r^\frac{1}{2}(n_1  + \textstyle\frac{d}{2}-\beta)-m
\bigr],\ \gamma^{\mu_1} ,\ \gamma^{\nu_1}
\Bigr)\Phi_{(\mu)_{n_1},\ (\nu)_{n_2}} \hspace{-0.5em}=
\Phi_{\{(\mu)_{n_1},\nu_1\}\nu_2...\nu_{n_2}}=0.\nonumber
\end{eqnarray}
The only peculiarities are as follows. Instead of case of HS
symmetry algebra, we need to consider the case of respective
superalgebra, and therefore the construction of VM and BFV-BRST
operator is more non-trivial, but exists (see for details
\cite{adsfermBKR,  Sakharov}). Because of presence in the total
Hilbert space $\mathcal{H}_{tot}$ the bosonic ghosts of any power
the stage of reducibility  grows with spin $(s_1,s_2)$. At last,
for the case of fermionic HS fields we must obtain Lagrangian
which is linear in derivatives $\nabla_{\mu}$. But prescription
which follows from bosonic case leads to second-order Lagrangian.
However, this problem are effectively overcome by means of partial
gauge fixing and solving of part of equations of motion. Doing so,
the LF for mixed-symmetry fermionic HS fields may be constructed
\cite{flatfermmix, Sakharov}.
\section{Summary}
We have briefly considered the method of constructing the LF for
free massive mixed-symmetry HS fields on AdS${}_d$ space in the
framework of  BFV-BRST approach. To do so, we have constructed new
auxiliary representation for non-linear algebra which serve for
conversion procedure of initial HS symmetry algebra. Then, we have
sketched a details of systematic way to find BFV-BRST operator for
non-linear operator algebra and presented a proper construction of
gauge LF basically for bosonic HS fields. The
Eqs.(\ref{LEoM}),({\ref{S}}) present the basic results of the
paper being the first step to interacting theory, following in
part to the research \cite{BuchbinderTsulaia}.

\paragraph{Acknowledgements} The author
thanks the organizers of the Bo\-go\-lyu\-bov'09 Conference (RAS -
JINR, Moscow - Dubna, Russia) for support and hospitality.
%%%%%%%%%%%%%%%%%%%%%%%%%%
%Plan:
%Introduction
%1. Bosonic  fields in AdS spaces
%2. Construction of auxiliary representations for non-linear algebras, its
% Fock space realizations and computer verification
%3. BRST operators for converted non-linear algebras
%4. Unconstrained Lagrangian formulations
%%5. Summary
%
%%%%%%%%%%%%%%%%%%%%%%%%%%%%%%%%%%%%%%%%%%%%%%%%%%%%%%%%%%%%%%%%%%%%

\end{document}